\documentclass[11pt,a4paper]{article}
\usepackage{fullpage}
\usepackage{authblk}
\usepackage{tikz}
\usepackage{ifthen}
\usepackage{graphics,epsfig}
\usepackage{amsmath,amsfonts,amssymb}
\usepackage{algorithm}
\usepackage{algpseudocode}
\usepackage{color}

\newtheorem{theorem}{Theorem}
\newtheorem{lemma}[theorem]{Lemma}

\newtheorem{definition}[theorem]{Definition}
\newcommand{\qed}{\hfill $\Box$ \medbreak}
\newenvironment{proof}{\noindent {\bf Proof.}}{\qed}

\newcommand{\accept}{\mbox{\sl accept}}
\newcommand{\reject}{\mbox{\sl reject}}
\newcommand{\rtable}{\mbox{\sf table}}
\newcommand{\rport}{\mbox{\sf port}}
\newcommand{\rname}{\mbox{\sf name}}
\newcommand{\certif}{\mbox{\sf certificate}}
\newcommand{\bunch}{\mbox{\sf bunch}}
\newcommand{\cluster}{\mbox{\sf cluster}}
\newcommand{\ball}{\mbox{\sf ball}}
\newcommand{\rnext}{\mbox{\sf next}}
\newcommand{\Dir}{\mbox{\sf Dir}}
\newcommand{\length}{\mbox{\rm length}}
\newcommand{\verif}{\mbox{\sc verif}}
\newcommand{\argmin}{\mbox{\rm argmin}}
\newcommand{\col}{\mbox{\sl color}}
\newcommand{\scol}{\mbox{\rm \scriptsize color}}
\title{Certification of Compact Low-Stretch Routing Schemes}

\author[1]{Alkida Balliu\thanks{Additional support from ANR Project DESCARTES. This work was done during the first author visit to Institut de Recherche en Informatique Fondamentale (IRIF) in 2016-2017.}}
\author[2]{Pierre Fraigniaud\thanks{Additional support from ANR Project DESCARTES and Inria Project GANG. }}

\affil[1]{Gran Sasso Science Institute, L'Aquila, Italy.}
\affil[2]{Institut de Recherche en Informatique Fondamentale\\CNRS and University Paris Diderot, France.}

\date{}

\begin{document}
\maketitle
\begin{abstract}
On the one hand, the correctness of routing protocols in networks is an issue of utmost importance for guaranteeing the 	delivery of messages from any source to any target. On the other hand, a large collection of \emph{routing schemes} have been proposed during the last two decades, with the objective of transmitting messages along short routes, while keeping the routing tables  small. Regrettably, all these schemes share the property that an adversary may modify the content of the routing tables with the objective of, e.g., blocking the delivery of messages between some pairs of nodes, without being detected by any node. 

In this paper, we present a simple \emph{certification} mechanism which enables the nodes to locally detect any alteration of their routing tables. In particular, we show how to locally verify the stretch-3 routing scheme by Thorup and Zwick [SPAA 2001] by adding certificates of $\widetilde{O}(\sqrt{n})$ bits at each node in $n$-node networks, that is, by keeping the memory size of the same order of magnitude as the original routing tables. We also propose a new \emph{name-independent} routing scheme using routing tables of size $\widetilde{O}(\sqrt{n})$ bits. This new routing scheme can be locally verified using certificates on $\widetilde{O}(\sqrt{n})$ bits. Its stretch is~3 if using handshaking, and~5 otherwise. 
\end{abstract}

\section{Introduction}

\subsection{Context}

A \emph{routing scheme}  is a mechanism enabling to deliver messages from any source to any target in a network. The latter is typically modeled as an undirected connected weighted graph $G=(V,E)$ where $V$ models the set of routers and $E$ models the set of communication links between routers. All edges incident to a degree-$d$ node are labeled from $1$ to~$d$, in an arbitrary manner, and the label at a node $u$ of an incident edge $e$  is called the \emph{port number} of edge $e$ at~$u$. A routing scheme consists of a way of assigning a routing \emph{table} to every node of the given network. These tables should contain enough information so that,  for every target node~$t$,  each node is able to compute the port number of the incident edge through which it should forward a message of destination~$t$. The routing tables must collectively guarantee that every message of any source $s$ and any target $t$ will eventually be delivered to $t$. 

Two scenarios are generally considered in the literature. One scenario allows the routing scheme to assign \emph{names} to the nodes, and each target is then identified by its given name. The other, called \emph{name independent}, is assuming that fixed names are given a priori (typically, a name is restricted to be the identity of a node), and the scheme cannot take benefit of naming nodes for facilitating routing. 

Among many criteria for evaluating the quality of routing schemes, including, e.g., the time complexity for constructing the routing tables, the two main parameters characterizing a routing scheme are the \emph{size} of its routing tables, and the \emph{stretch}. The stretch of a routing scheme is the maximum, taken over all pairs of source-target nodes, of the ratio between the length of the route generated by the scheme from the source to the target, and the length of a shortest path between these two nodes. During the last two decades, there has been an enormous effort to design \emph{compact} routing scheme (i.e., schemes using small tables) of \emph{low} stretch (i.e., with stretch upper bounded by a constant) -- see, e.g., \cite{AGM06,AGMNT04,ACLRT03,ABLP89,FG01,GG01,PU88,SK85,TZ01}. A breakthrough was achieved in~\cite{TZ01} where almost tight tradeoffs between size and stretch were explicitly demonstrated.  In particular, \cite{TZ01} showed how to design a routing scheme with tables of size $\widetilde{O}(\sqrt{n})$ bits and stretch~3, in any network\footnote{\small The notations $\widetilde{O}$ and $\widetilde{\Omega}$ ignore polylogarithmic factors.}. 

All the aforementioned routing schemes share the property that nodes do not have the capability to realize that the routing tables have been modified (either involuntarily or by an attacker). That is, a group of nodes may be provided with routing tables which look consistent with a desired routing scheme, but which do not achieve the desired performances of that scheme (e.g., large stretch, presence of loops, etc.). Indeed, the nodes are not provided with sufficient information to detect such an issue locally, that is, by having each node inspecting only the  network structure and the tables assigned to nodes in its vicinity. 

\subsection{Objective}

The objective of this paper is, given a routing scheme, to design a mechanism enabling each node to locally detect the presence of falsified routing tables, in the following sense. If one or more tables are erroneous, then at least one node must be able to detect that error by running a verification algorithm exchanging messages only between neighboring nodes. 

Our mechanism for locally verifying the correctness of routing tables is inspired from proof-labeling schemes~\cite{KKP10}. It is indeed based on assigning to each node a \emph{certificate}, together with its routing table, and designing a distributed verification algorithm that checks the consistency of these certificates and tables by having each node inspecting only its certificate and its routing table, and the certificate and routing table of each of its neighbors. The set of certificates assigned to the nodes and the verification algorithm running at all nodes in parallel must satisfy that: (1) if all tables are correctly set, then, with some appropriate certificates, all nodes \emph{accept}, and (2) if one or more tables are incorrectly set, then, for every assignment of the certificates, at least one node must \emph{reject}. The second condition guarantees that the verification algorithm cannot be cheated: if the tables are incorrect, there are no ways of assigning the certificates such that all nodes accept. 

Rephrasing the objective of the paper, our goal is to assign certificates to nodes, of size not exceeding the size of the routing tables, enabling the nodes to collectively verify the correctness of the routing tables, by having each node interacting with its neighbors only. 

\subsection{Our Results}

We show how to locally verify the stretch-3 size-$\widetilde{O}(\sqrt{n})$ routing scheme by Thorup and Zwick~\cite{TZ01}. Our certification mechanism uses certificates of $\widetilde{O}(\sqrt{n})$ bits at each node, that is, these certificates have size of the same order of magnitude as the original routing tables. Hence, verifying the scheme in~\cite{TZ01} can be done without modifying the scheme, and without increasing the memory space consumed by that scheme. We also show that the same holds for the whole hierarchy of routing schemes proposed in~\cite{TZ01} for providing a tradeoff between size and stretch. 

The situation appears to be radically different for name-independent routing schemes. The stretch-$3$ name-indepen\-dent routing scheme by Abraham et al.~\cite{AGMNT04} also uses tables of size $\widetilde{O}(\sqrt{n})$ bits. However, each table includes references to far away nodes, whose validity does not appear to be locally verifiable using certificates of reasonable size. On the other hand, a simplified version of the scheme in~\cite{AGMNT04} can be verified locally with certificates of size  $\widetilde{O}(\sqrt{n})$ bits, but its stretch becomes at least~7. Therefore, we propose a new name-independent routing scheme, with tables of size $\widetilde{O}(\sqrt{n})$ bits that can be verified using certificates on $\widetilde{O}(\sqrt{n})$ bits as well. This new routing scheme has stretch at most~5, and the stretch can even be reduced to~3 using handshaking\footnote{\small The handshaking mechanism is similar to DNS lookup in TCP/IP. It allows querying some node(s) for getting additional information about the route to the target.}. The routing scheme of Arias et al.~\cite{ACLRT03} has also stretch~5, but it does not appear to be locally verifiable with certificates of reasonable size, and using handshaking does not enable to reduce the stretch. 

All our results are summarized in Table~\ref{tab:summary}. 

\begin{table}[htp]
	\begin{center}\small 
		\begin{tabular}{c|c|c|c|c|}
			
			&  &  name &  &  \\
			scheme                  	&    stretch  &   independent & verifiable & comment \\ \hline
			\cite{TZ01}    	&  3 & no & \textbf{yes} & --   \\
			\cite{ACLRT03} &  5 & yes & \textbf{?} & -- \\
			\cite{AGMNT04} &  3 & yes & \textbf{?} & -- \\
			\textbf{this paper}		& \textbf{5} & \textbf{yes} & \textbf{yes} & -- \\
			\textbf{this paper}		&  \textbf{3} & \textbf{yes} & \textbf{yes} & handshaking \\
			\hline
		\end{tabular}
	\end{center}
	\caption{Summary of our results compared to previous work. All the listed routing schemes have space complexity of $\widetilde{O}(\sqrt{n})$ bits. Our verification algorithms for these schemes  use certificates on $\widetilde{O}(\sqrt{n})$ bits.}
	\label{tab:summary}
\end{table}

\subsection{Related Work}

The design of \emph{compact} routing tables, and the explicit identification of tradeoffs between the table size and the routes length was initiated thirty years ago, with the seminal work in~\cite{SK85} and \cite{PU88}. Since then, a large amount of papers were published on this topic, aiming at refining these tradeoffs, and at improving different aspects of the routing schemes, including routing in specific classes of graphs (see~\cite{G01,GP03}). In particular, routing schemes were designed for trees in~\cite{FG01,TZ01}, with space complexity $O(\log^2n/\log\log n)$ bits\footnote{\small The space complexity can be reduced to $O(\log n)$ if the designer of the routing scheme is also allowed to assign the port numbers to the nodes.}. This space complexity was shown to be optimal in~\cite{FG02}. 

It was proved~\cite{GP96} that,  in $n$-node networks, any shortest path routing scheme requires tables of size $\widetilde{\Omega}(n)$ bits. The aforementioned routing scheme in~\cite{TZ01} with stretch $3$  and space complexity  $\widetilde{O}(\sqrt{n})$ bits was shown to be optimal in~\cite{GG01}, in the following sense:  no routing scheme with space complexity $o(n)$ bits can achieve a stretch $s<3$, and, assuming the correctness of a conjecture by Erd\H{o}s regarding a tradeoff between girth and edge density in graphs, every routing scheme with stretch $s<5$ has space complexity $\Omega(\sqrt{n})$ bits. On the positive side,  \cite{TZ01} tightens the size-stretch tradeoff of \cite{PU88} by showing that, for every $k\geq 2$,  there exists a routing scheme with stretch $4k-5$ and space complexity $\widetilde{O}(n^{1/k})$ bits. (The stretch can be reduced to $2k-1$ using handshaking). Recently,  \cite{C13}~showed that, for $k\geq 4$,  a stretch $s=\alpha \,k$ with $\alpha<4$ can be achieved using routing tables of size $\widetilde{O}(n^{1/k})$. 

The distinction between name-independent routing sche\-mes, and routing schemes assigning specific names to the nodes was first made in~\cite{ABLP89}. Then, \cite{AP90}~presented  techniques for designing name-independent routing schemes with constant stretch and space complexity $o(n)$ bits. Almost 15 years after, \cite{ACLRT03} described a name-independent routing scheme with stretch~$5$ and space complexity $\widetilde{O}(\sqrt{n})$ bits. This was further improved in \cite{AGMNT04} thanks to a name-independent routing scheme with stretch~$3$ and space complexity $\widetilde{O}(\sqrt{n})$ bits. A couple of years later, \cite{AGM06} showed that there are tradeoffs between stretch and space complexity for name-independent routing schemes as well. Specifically,  \cite{AGM06} showed that, for any $k\geq 1$, there exists a  name-independent routing scheme with space complexity $\widetilde{O}(n^{1/k})$ bits and stretch $O(k)$.

The certification mechanism used in this paper is based on the notion of proof-labeling scheme introduced in~\cite{KKP10} in which an oracle, called \emph{prover}, assigns certificates to the nodes, and a distributed algorithm, called \emph{verifier}, checks that this certificates collectively form a proof that the global state of the network is legal with respect to a given boolean network  predicate. Proof-labeling schemes have been widely used in literature. For example, \cite{SS13} uses them to verify spanning trees in networks. This result has been extended in \cite{FLSW16}, where proof-labeling schemes are used to verify spanning trees in evolving networks that are evolving with time. Variants of proof-labeling schemes have been considered in, e.g., \cite{BFP-S15,FKP13}, and \cite{GS16}. More generally, see~\cite{FF16} for a survey of distributed decision. 

\subsection{Structure of the Paper}

In Section~\ref{sec:definitions}, we recall the main concepts related to the design of routing schemes, and those related to distributed verification. Section~\ref{sec:namedependent} is dedicated to routing schemes assigning names to nodes:  we briefly recall the main features of the scheme in~\cite{TZ01}, and then show how to locally verify that scheme using small certificates. In Section~\ref{sec:unlikely}, we show that the name-independent routing scheme in~\cite{AGMNT04} is unlikely to be locally verifiable with small certificates. Therefore, Section~\ref{sec:newnameind} presents a new name-independent routing scheme, that is shown to be locally verifiable with small certificates. This new scheme has space-complexity $\widetilde{O}(\sqrt{n})$ bits, but stretch~5. Section~\ref{sec:another} shows that the stretch-5 scheme in~\cite{ACLRT03} is, like the one in~\cite{AGMNT04},  unlikely to be locally verifiable with small certificates. Finally, Section~\ref{sec:conclusion} concludes the paper with pointers to research directions worth to be investigated. 

\section{Definitions}
\label{sec:definitions}

\subsection{Routing Schemes}

Let $\cal{F}$ be a family of edge-weighted graphs with edges labeled at each node by distinct port numbers from~1 to the degree of the node. The weights are all positive, and the weight of edge $e$ represents its length. It is thus denoted by $\length(e)$. The nodes are given distinct identities stored on $O(\log n)$ bits. For the sake of simplifying notations, we do not make a distinction between a node $v$ and its identity, also denoted by~$v$. Given two nodes $u,v$, we denote by 
\[
\delta (u,v) = \mbox{weighted distance between $u$ and $v$}. 
\]
Given an edge $e$ of extremity $u$, the port number of $e$ at $u$ is denoted by $\rport_u(e)$. 

A \emph{routing scheme} for $\cal{F}$ is a mechanism assigning a \emph{name}, $\rname(u)$, and a routing \emph{table}, $\rtable(u)$, to every node $u$ of every graph $G\in\cal{F}$ such that, for any pair $(s,t)$ of nodes of any $G\in\cal{F}$, there exists a path $u_0,u_1,\dots,u_k$ from $s$ to $t$ in $G$ with $u_0=s$, $u_k=t$, and
\begin{equation}\label{eq:routing}
	\rtable(u_i)\big (\rname(t)\big )=\rport_{u_i}(\{u_i,u_{i+1}\})
\end{equation}
for every $i=0,\dots,k-1$. That is, every intermediate node~$u_i$, $0\leq i < k$, can determine on which of its ports the message has to be forwarded, based solely on its routing table, and on the name of the target. In Eq.~\ref{eq:routing}, each table is viewed as a function taking names as arguments, and returning port numbers. The path $u_0,u_1,\dots,u_k$ is then called the \emph{route} generated by the scheme from~$s$ to~$t$. It is worth pointing out the following observations. 

\medskip

\noindent $\bullet$ \emph{Name-independent} routing schemes are restricted to use names that are fixed a priori, that is, the name of a node is its identity, i.e., $\rname(v)=v$ for every node~$v$. Instead, name-dependent routing schemes allow names to be  set for facilitating routing, and names are typically just bounded to be storable on a polylogarithmic number of bits. 

\medskip 

\noindent $\bullet$ The \emph{header} of a message is the part of that message containing all information enabling its routing throughout the network. The header of a message with destination $t$ is typically $\rname(t)$. However, some routing schemes ask for message headers that can be modified. This holds for both name-dependent schemes (like, e.g., \cite{TZ01}) and name-independent schemes (like, e.g., \cite{AGMNT04}). The typical scenario requiring  modifiable headers is when a message is routed from source $s$ to target $t$ as follows. From $\rname(t)$ and $\rtable(s)$, node $s$ can derive the existence of some node $v$ containing additional  information about how to reach $t$. Then the message is first routed from $s$ to $v$, and then from $v$ to $t$. Distinguishing these two distinct parts of the routes from $s$ to $t$ often requires to use different headers. In case of modifiable headers, Eq.~\eqref{eq:routing} should be tuned accordingly as the argument of routing is not necessarily just a name, but a header. 

\medskip 

\noindent $\bullet$ Some routing schemes may use a mechanism called \emph{handshaking}, which is an abstraction of mechanisms such as  Domain Name System (DNS) enabling to recover an IP address from its domain name. Let us consider the aforementioned scenario where a routing scheme routes a message from $s$ to $t$ via an intermediate node $v$ identified by $s$ from $\rname(t)$ and $\rtable(s)$. One can then enhance the routing scheme by a handshaking mechanism, enabling $s$ to query $v$ directly, and to recover the information stored at $v$ about $t$ at no cost. Then $s$ can route the message directly from $s$ to $t$, avoiding the detour to $v$. Handshaking is used in~\cite{TZ01} to reduce the stretch of routing schemes with space complexity $\widetilde{O}(n^{1/k})$ bits from $4k-5$ to $2k-1$, for every $k>2$. 

\medskip

The \emph{size} of a routing scheme is the maximum, taken over all nodes $u$ of all graphs in $\cal{F}$, of the memory space required to encode the function $\rtable(u)$ at  node~$u$. 

The \emph{stretch} of a routing scheme is the maximum, taken over all pairs $(s,t)$ of nodes in all graphs $G\in \cal{F}$, of the ratio of the length of the route from $s$ to $t$ (i.e., the sum of the edge weights along that route) with the weighted distance between $s$ and $t$ in~$G$. 

\subsection{Distributed Verification}
\label{subsec:certification}

Given a graph $G$, a \emph{certificate function} for $G$ is a function 
\[
\certif : V(G)\to\{0,1\}^*
\] 
assigning a certificate, $\certif(u)$, to every node $u\in V(G)$. A \emph{verification} algorithm is a distributed algorithm running concurrently at all nodes in parallel. At every node $u\in V(G)$ of every graph $G\in \cal{F}$, the algorithm takes as input $\certif(u)$ and $\rtable(u)$ assigned to node $u$, as well as the collection of pairs $(\rtable(v),\certif(v))$ assigned to all neighbors $v$ of $u$, and outputs \accept\/ or \reject. 

\begin{definition}
	A routing scheme for $\cal{F}$ is \emph{verifiable} if there exists a verification algorithm $\verif$ such that, for every $G\in \cal{F}$, 
	\begin{itemize}
		\item if the tables given to all nodes of $G$ are the ones specified by the routing scheme, then there exists a certificate function  for $G$ such that the verification algorithm  $\verif$ outputs \accept\/ at all nodes; 
		\item if some tables given to some nodes of $G$ differ from the ones specified by the routing scheme, then, for every certificate function for $G$, the verification algorithm  $\verif$ outputs reject in at least one node.  
	\end{itemize}
\end{definition}

The second bullet guarantees that if an adversary modifies some routing tables, or even just a single bit of a single table, then there are no ways it can also modify some, or even all certificates so that to force all nodes to accept: at least one node will detect the change. Of course, this node does not need to be the same for different modifications of the routing tables, or for different certificates. 

\paragraph{Remark.} The above definition is the classical definition of proof-labeling scheme applied to verifying routing schemes. In particular, it may well be the case that a correct labeling scheme be rejected if the certificates have not been set appropriately, just like a spanning tree $T$ will be rejected by a proof-labeling scheme for spanning trees if the certificates have been set for another spanning tree $T'\neq T$. 

\section{Name-dependent routing scheme}
\label{sec:namedependent}

In this section, we show how to verify the stretch-$3$ routing scheme by Thorup and Zwick in \cite{TZ01}. This scheme uses tables of $\widetilde{O}(\sqrt{n})$ bits of memory at each node. We show the following: 

\begin{theorem}\label{theo:classic}
	The stretch-$3$ routing scheme by Thorup and Zwick in \cite{TZ01} can be locally verified using certificates of size $\widetilde{O}(\sqrt{n})$ bits.
\end{theorem}

Before proving the theorem, we first recall the general structure of the routing scheme in~\cite{TZ01}. 

\subsection{The routing scheme}
\label{subsec:TZsummary}

The routing scheme assigns  names and  tables to the nodes of  every $G=(V,E)$ as follows. 

\subsubsection{Landmarks, Bunch and Clusters}

The routing scheme in~\cite{TZ01} uses the notion of \emph{landmarks} (a.k.a.~\emph{centers} in the ``older'' terminology of \cite{TZ01}). These landmarks form a subset $L\subseteq V$ of nodes. For $v\in V$, let $l_v$ denote the landmark closest to $v$ in $G$. For every $v\in V$, the \emph{bunch} of $v$ with respect to the set $L$ is defined as follows. 
\[
\bunch(v) = \{u\in V : \delta(v,u)<\delta(l_v, v) \}.
\]
%
%
%
The routing scheme in~\cite{TZ01} also uses the notion of  \emph{cluster}. For every node $v\in V$, 
\[
\cluster(v) = \{u\in V : \delta(v,u)<\delta(l_u, u) \}.
\] 
As a consequence, for every $u,v \in V$, we have 
\[
u \in \cluster(v) \iff v\in \bunch(u).
\]
Note that since, for every  $v\in V$, and every  $l \in L$, we have $l\notin \bunch(v)$, it follows that $\cluster(l)=\emptyset$ for every $l\in L$. By construction of the bunches and the clusters, it also holds that, for every  $v\in V$, $\cluster(v)\cap L=\emptyset$. Also, clusters satisfy the following property.

\begin{lemma}[\cite{TZ01}] \label{propCluster}
	If $u\in \cluster(v)$ then, for every node  $w$ on a shortest path between~$u$ and~$v$, we have $u\in \cluster(w)$ 
\end{lemma}


In \cite{TZ01}, the landmarks are chosen by an algorithm that samples them uniformly at random in $V$ until the following holds: for every node $v$,  $|\cluster(v)| < 4\sqrt{n}$. It is proved that this algorithm returns, w.h.p., a set of landmarks of size at most $2 \log(n)\sqrt{n}$.

\subsubsection{Names and tables}

For every two nodes $v$ and $t$, let 
\[
\rnext(v,t)
\]
be the port number of an edge incident to $v$ on a shortest path between $v$ and $t$. Each node $t\in V$ is assigned a $ 3\lceil \log(n) \rceil $-bit name 
as follows:
\[
\rname(t) = (t, l_t, \rnext(l_t,t)).
\]
Each node $v\in V$ then stores the following information in  its routing table, $\rtable(v)$: 
\begin{itemize}
	\item the identities of all the landmarks $l\in L$;
	\item the identities of all nodes  $t\in\cluster(v)$;
	\item the set $\{\rnext(v,t) : t\in L \cup \cluster(v)\}$.  
\end{itemize}
All these information can be stored using $\widetilde{O}(\sqrt{n})$ bits.

\subsubsection{Routing}

Note that, by Lemma~\ref{propCluster}, any message from a node $v$ to a node in $\cluster (v)$ reaches its target along a shortest path. The same holds for landmarks since every node is given information about how to reach every landmark. In general, let us assume that $v$ wants to send a message to some node $t$ with label $(t, l_t, \rnext(l_t,t))$ that is neither a landmark nor belongs to $\cluster(v)$. In this case, $v$ extracts the landmark  $l_t$ nearest to $t$ from $t$'s name (note here the impact of allowing the scheme to assign specific names to nodes), and forwards the message through the port on a shortest path towards $l_t$ using the information $v$ has in its table. Upon reception of the message, $l_t$ forwards the message towards $t$ on a shortest path using $\rnext(l_t,t)$ (this information can also be extracted from the name of $t$), to reach a node $z\in \bunch(t)$. At last, since $z\in \bunch(t)$, we have $t\in \cluster(z)$, which means that $z$ can route to $t$ via a shortest path using the information available in its table. By Lemma~\ref{propCluster}, this also holds for every node along a shortest path between $z$ and $t$. Using symmetry, and triangle inequality, \cite{TZ01} shows that this routing scheme guarantees stretch 3.

\subsection{Proof of Theorem~\ref{theo:classic}}

In order to enable local verification of the stretch-3 routing scheme in~\cite{TZ01}, a certificate of size $\widetilde{O}(\sqrt{n})$ bits is given to each node. Let $G=(V,E)$ be an  undirected graph with positive weights assigned to its edges, and a correct assignment of the routing tables to the nodes according to the specifications of~\cite{TZ01} as summarized in Section~\ref{subsec:TZsummary}. Then each node $v\in V$ is assigned a certificate composed of: 
\begin{itemize}
	\item the distance between $v$ and every landmark in $L$;
	\item the distance between $v$ and every node in $\cluster(v)$;
	\item the set $\{\delta(t,l_t) : t\in \cluster(v)\}$.
\end{itemize}
As claimed, all these information can be stored using $\widetilde{O}(\sqrt{n})$ bits of memory.

We assume, without loss of generality, that all nodes know~$n$ (verifying the value of $n$ is easy using a proof-labeling scheme with $O(\log n)$-bit certificates \cite{KKP10}). The verification of the routing scheme then proceeds as follows. We describe the verification algorithm $\verif$ running at node $v\in V$. This verification goes in a sequence of  steps. At each step, either $v$ outputs \reject\/ and stops, or it goes to the next step. 

We denote by $L(v)$, $C(v)$, and $\{N(v,t) : t\in L(v) \cup C(v)\}$ the content of the routing table of $v$. These entries are supposed to be the set of landmarks, the cluster of $v$, and the set of $\rnext$-pointers given to~$v$, respectively. We also denote by $d$ the distance given in the certificates. That is, node $v$ is given a set $\{d(v,t) : t\in L(v)\cup C(v)\}$ and a set $\{d(t,l_t) : t\in C(v)\}$ where $l_t$ is supposed to be the node in $L(v)$ closest to $t\in C(v)$. Of course, if a node does not have a table and a certificate of these forms, then it outputs \reject. So, we assume that all tables and certificates have the above formats. The algorithm proceeds as follows at every node~$v$. Node~$v$ checks that

\begin{enumerate} 
	\item \label{item:size}   the information in its table satisfy $|C(v)|\le 4 \sqrt{n}$ and $|L(v)|\le 2\log n \sqrt{n}$ bits; 

	\item  \label{item:sameL}   it has the same set of landmarks as its neighbors; 

	\item  \label{item:distL}  for every $l \in L(v)$, there exists a neighbor $u$ of $v$ satisfying 
	\[
	d(v,l)=\length(\{v,u\}) + d(u,l),
	\]
	and all other neighbors satisfy
	\[
	d(v,l) \leq \length(\{v,u\}) + d(u,l),
	\]
	with $N(v,l)$ pointing to a neighbor $u$ satisfying $d(v,l)=\length(\{v,u\}) + d(u,l)$;

	\item \label{item:clusterlandmark}  if $v\in L(v)$ then $C(v)=\emptyset$;

	\item  \label{item:distC}   $v\in C(v)$ and, for every node $t \in C(v)$, there exists a neighbor $u$ of $v$ satisfying $t\in C(u)$ with 
	\[
	d(v,t)=\length(\{v,u\}) + d(u,t),
	\]
	and every neighbor $u$ of $v$ with $t\in C(u)$ satisfies 
	\[
	d(v,t) \leq \length(\{v,u\}) + d(u,t),
	\]
	with $N(v,t)$ pointing to a neighbor $u$ satisfying $t\in C(u)$ and $d(v,t)=\length(\{v,u\}) + d(u,t)$;
	\item \label{item:additional} for every node $t \in C(v)$, for every neighbor $u$ of $v$ satisfying $t \in C(u)$, the distance $d(t,l_t)$ in the certificate of $u$ is equal to the distance $d(t,l_t)$ in the certificate of~$v$;

	\item  \label{item:cluster} for every $t\in C(v)$, it holds that $d(v,t)<d(t,l_t)$;

	\item  \label{item:bunch}  for every neighbor $u$, and every $t\in \cluster(u)\setminus \cluster(v)$, it holds that 
	\[
	\length\{(v,u)\} + d(u,t) \ge d(t,l_t).
	\]
	
\end{enumerate}

\noindent If $v$ passes all the above tests, then $v$ outputs \accept, else it outputs \reject. 

\medskip 

We now establish the correctness of this local verification algorithm, that is, we show that it satisfies the specification stated in Section~\ref{subsec:certification}. First, by construction, if all tables are set according to~\cite{TZ01}, that is, if, for every node $v$, $L(v)=L$, $C(v)=\cluster(v)$ and $N(v,t)=\rnext(v,t)$ for all $t\in L \cup \cluster(v)$, then every node running the verification algorithm with the appropriate certificate  $\{\delta(v,t) : t\in L \cup \cluster(v)\} \cup \{\delta(t,l_t) : t\in \cluster(v)\}$ where $l_t$ is the node in $L$ closest to $t\in \cluster(v)$, will face no inconsistencies with its neighbors, i.e., all the above tests are passed, leading every node to accept, as desired. 

So, it remains to show that if some tables are not what they should be according to~\cite{TZ01}, then, no matter the certificates assigned to the nodes, at least one node will fail one of the tests. 

If all nodes output \accept, then, by Step~\ref{item:size}, all routing tables are of the appropriate size.  Also, by Step~\ref{item:sameL}, the set $L$ of landmarks given to the nodes is the same for all nodes, as otherwise there will be two neighbors that would have different sets. Moreover, by Step~\ref{item:distL}, we get that, at every node $v$, the distances of this node to the landmarks, as stored in its certificate, are correct, from which we infer that $N(v,l)$  is appropriately set in the table of~$v$, that is $N(v,l)=\rnext(v,l)$ for every $l\in L$. Hence, if all the tests in Steps~\ref{item:size}-\ref{item:distL} are passed, all the data referring to $L$ in both the tables and the certificates are consistent. In particular, every node $v$ knows the landmark $l_v$ which is closest to it, and its distance $\delta(v,l_v)$. 

We now show that if all these tests as well as the remaining tests are passed, then the clusters in the tables are correct, w.r.t.~$L$, as well as the $\rnext$-pointers.  We first show that, if  all tests are passed, then, for every node $v\in V$, 
\[
C(v)\subseteq \cluster(v).
\] 
By Step~\ref{item:clusterlandmark}, this latter equality holds for every landmark~$v$.  By Step~\ref{item:distC}, we get that, at every node $v$, the distance of this node to every node $t\in C(u)$, as stored in its certificate, are correct, from which we infer that $N(v,t)$  is appropriately set in the table of~$v$, that is $N(v,t)=\rnext(v,t)$ for every $t\in C(v)$. By Step~\ref{item:additional}, we get that, for every $t\in C(v)$, we do have $d(l_t,t)=\delta(l_t,t)$. Indeed, this equality will be checked by all nodes on a shortest path between $v$ and $t$ (whose existence is guaranteed by Step~\ref{item:distC}), and $t$ has the right distance $\delta(t,l_t)$ in its certificate by Step~\ref{item:distL}. Recall that 
\[
\cluster(v) = \{t\in V ~|~ \delta(v,t)<\delta(l_t, t) \}
\] 
where $l_t$ is the landmark closest to~$t$. Step~\ref{item:cluster} precisely checks that inequality. 

It remains to show that there are no nodes in $\cluster(v)$ that are not in $C(v)$. Assume that there exists $t\in \cluster(v) \setminus C(v)$, and let $P$ be a shortest path between $v$ and $t$. Let $v'$ be the closest node to $v$ on $P$ such that $t\in C(v')\subseteq \cluster(v')$. Note that such a node $v'$ exists as $t\in C(t)$. Let $v''$ be the node just before $v'$ on $P$ traversed from $v$ to $t$. By Lemma~\ref{propCluster}, since $t\in \cluster(v)$, we also have $t\in \cluster(v'')$. We have $t\in \cluster(v'') \setminus C(v'')$. Therefore $\delta(v'',t)<\delta(l_t,t)$. Now, 
\[
\delta(v'',t)=\length(\{(v'',v'\})+\delta(v',t)
\] 
because $P$ is a shortest path between $v''$ and $t$ passing through~$v'$. So, 
\[
\length(\{(v'',v'\})+\delta(v',t)<\delta(l_t,t).
\] 
Step~\ref{item:bunch} guarantees that it is not the case. Therefore, there are no nodes in $\cluster(v) \setminus C(v)$. It follows that $\cluster(v)=C(v)$ for all nodes~$v$. This completes the proof of Theorem~\ref{theo:classic}.

\section{Name-independent Routing Scheme}
\label{sec:unlikely}

The purpose of this section is twofold. First, it serves as recalling basic notions that will be helpful for the design of our new name-independent routing scheme. Second, it is used to show why the known name-independent routing scheme in \cite{AGMNT04} appears to be difficult, and perhaps even impossible to verify locally. 

\subsection{The Routing Scheme of \cite{AGMNT04}}

The stretch-3 name-independent routing scheme of \cite{AGMNT04} uses $\widetilde{O}(\sqrt{n})$ space at each node. We provide a high level description of that scheme. Recall that, in name-independent routing, a target node is referred only by its identity. That is, $\rname(t)$ is the identity of $t$, i.e., $\rname(t)=t$. 

Let $G=(V,E)$. For every node $v\in V$, the \emph{vicinity ball} of $v$, denoted by $\ball(v)$, is the set of the $4\lceil \alpha \log(n) \sqrt{n} \rceil$ closest nodes to $v$, for a large enough constant $\alpha>0$, where ties are broken using the order of node identities. By this definition, if $u\in \ball(v)$ and $w$ is on a shortest path between $v$ and $u$, then $u\in \ball(w)$.

\subsubsection{Color-Sets}

In  \cite{AGMNT04}, the nodes are partitioned into sets $C_1,\dots C_{\sqrt{n}}$, called \emph{color-sets}, and, for $i=1,\dots,\sqrt{n}$, the nodes in color-set $C_i$ are assigned the same color $i$. For every node $v\in V$, its vicinity ball, $\ball(v)$, contains at least one node from each color-set. To get this,  the color of a given node $v$ is determined by a hash function $\col$ which, given the identity of a node, maps that identity to a color in $\{1,\dots,\sqrt{n}\}$. This mapping from identities to colors is balanced in the sense that at most $O(\log n \sqrt{n})$ nodes map to the same color. A color is chosen arbitrarily, and all nodes with that color are considered to be the landmarks. Let $L$ be the set of landmarks. It holds that $|L|=O(\log n \sqrt{n})$. Also, each node $v\in V$ has at least one landmark in its vicinity ball. We fix, for each vicinity ball $\ball(v)$, an arbitrary landmark, denoted by $l_v$. 

\subsubsection{Routing in trees} 

This construction in \cite{AGMNT04} makes use of routing schemes in trees. More precisely, they use the results from \cite{FG01,TZ01}, which states that there exists a shortest-path (name-dependent) routing scheme for trees using names and tables both on $O(\log^2(n)/\log \log (n))$ bits in $n$-node trees. For a tree $T$ containing node $v$, let $\rtable_{T}(v)$ and $\rname_T(v)$ denote the routing table of node $v$ in $T$, and the name of $v$ in $T$, as assigned by the scheme in  \cite{FG01}.  

\subsubsection{The routing tables}
\label{subsubsec:trt}

For any node $v$, let $T(v)$ be a shortest-path spanning tree rooted at $v$. Let $P(v,w,u)$ be a path between $v$ and $u$ composed of a shortest path between $v$ and $w$ and a shortest path between $w$ and $u$. Such a path is said to be \emph{good} for $(v,u)$ if $v \in \ball(w)$, and there exists an edge $\{x,y\}$ along a shortest path between $w$ and $u$ with $x\in \ball(w)$ and  $y\in \ball(u)$;

Every node $v\in V$ stores the following information in its routing table $\rtable(v)$:

\begin{itemize}
	
	\item the hash function $\col$ that maps identities in colors; 
	
	\item the identity of every node $u\in\ball(v)$, and the port number $\rnext(v,u)$;  
	
	\item for every landmark $l\in L$, the routing table $\rtable_{T(l)}(v)$ for routing in $T(l)$;
	
	\item for every node $u\in \ball(v)$, the routing table $\rtable_{T(u)}(v)$ for routing in $T(u)$; 
	
	\item the identities of all nodes with same color as $v$, and, for each such node $u$, the following additional information: 
	
	\begin{itemize}
		
		\item  if there are no good paths $P(v,w,u)$, then $v$ stores 
		\begin{equation}\label{eq:caseA}
			\big ( \rname_{T(l_u)}(l_u), \rname_{T(l_u)}(u) \big ).
		\end{equation}
		
		\item if there exists a good path $P(v,w,u)$, then let us pick a good path $P$ of minimum length among all good paths; then, let us compare its length $|P|$ with the length $|Q|$ of the path $Q$ composed of a shortest path between $v$ and $l_u$, and a shortest path between $l_u$ and $u$; provide $v$ with  
		\begin{equation*}
			\big ( \rname_{T(l_u)}(l_u), \rname_{T(l_u)}(u) \big )
		\end{equation*}
		if $|Q|\leq |P|$, and with 
		\begin{equation}\label{eq:caseB2}
			\big ( \rname_{T(v)}(w), x, \rport_x(\{x,y\}), \rname_{T(y)}(u)\big)
		\end{equation}
		otherwise.
		
	\end{itemize}
\end{itemize}

Storing the hash function $\col$ requires $O(\sqrt{n})$ bits. $L$~and $\ball(v)$ are both of size $\widetilde{O}(\sqrt{n})$ bits. Moreover, the number of nodes with identical color is $\widetilde{O}(\sqrt{n})$, for every color. Finally, shortest-path routing in any tree can be achieved using tables and names of size $O(\log^2 n/\log\log n)$ bits~\cite{FG01,TZ01}. It follows that $|\rtable(v)|= \widetilde{O}(\sqrt{n})$, as desired. 

\subsubsection{Routing }

Routing from a source $s$ to a target $t$ is achieved in the following way (see~\cite{AGMNT04} for more details). If $t\in \ball(s)$, or $t\in L$, or $s$ and $t$ have the same color, then $s$ routes to $t$ using the information available in its table. More specifically, if $t\in \ball(s)$ or $t\in L$, then $s$ sets the header of the message as just the identity of the target~$t$. Instead, if  $s$ and $t$ have the same color but $t\notin \ball(s) \cup L$, the source $s$ sets the header as one of the two possible cases presented in Eq.~\eqref{eq:caseA} and~\eqref{eq:caseB2}.  Otherwise, that is, if $t\notin \ball(s) \cup L$ and $\col(s)\neq\col(t)$, node $s$ routes the message towards some node $w\in \ball(s)$ sharing the same color as $t$. (The color of $t$ can be obtained by hashing the identity of~$t$). The header is set to $t$, and $w$ will change the header upon reception of this message, according to the rules previously specified. It is shown in \cite{AGMNT04} that this routing guarantees a stretch~3.

\subsection{On the difficulty of locally verifying the scheme in~\cite{AGMNT04}}

\begin{figure}[htb]
	\centering
	\includegraphics[width=0.7\textwidth]{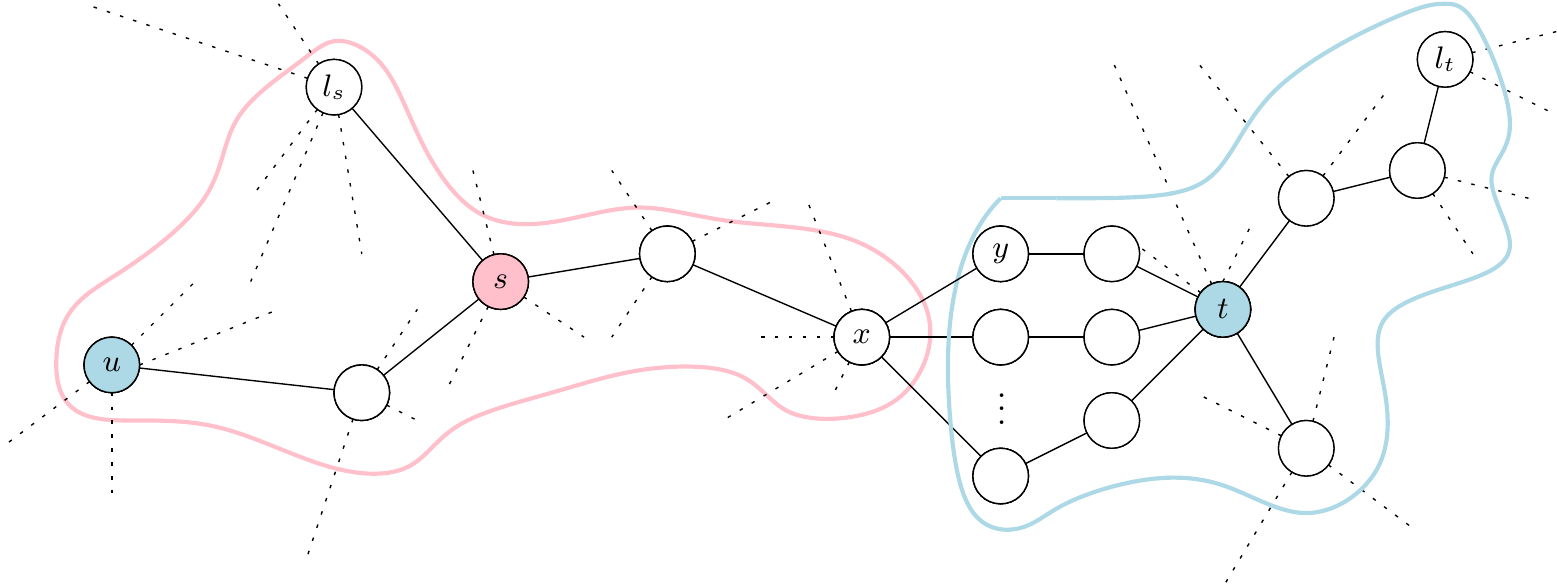}
	\caption{All the nodes in a shortest path between $s$ and $t$ belong to $\ball(s) \cup \ball(t)$.}
	\label{intersectBalls}
\end{figure}

\begin{figure}[htb]
	\includegraphics[width=0.7\textwidth]{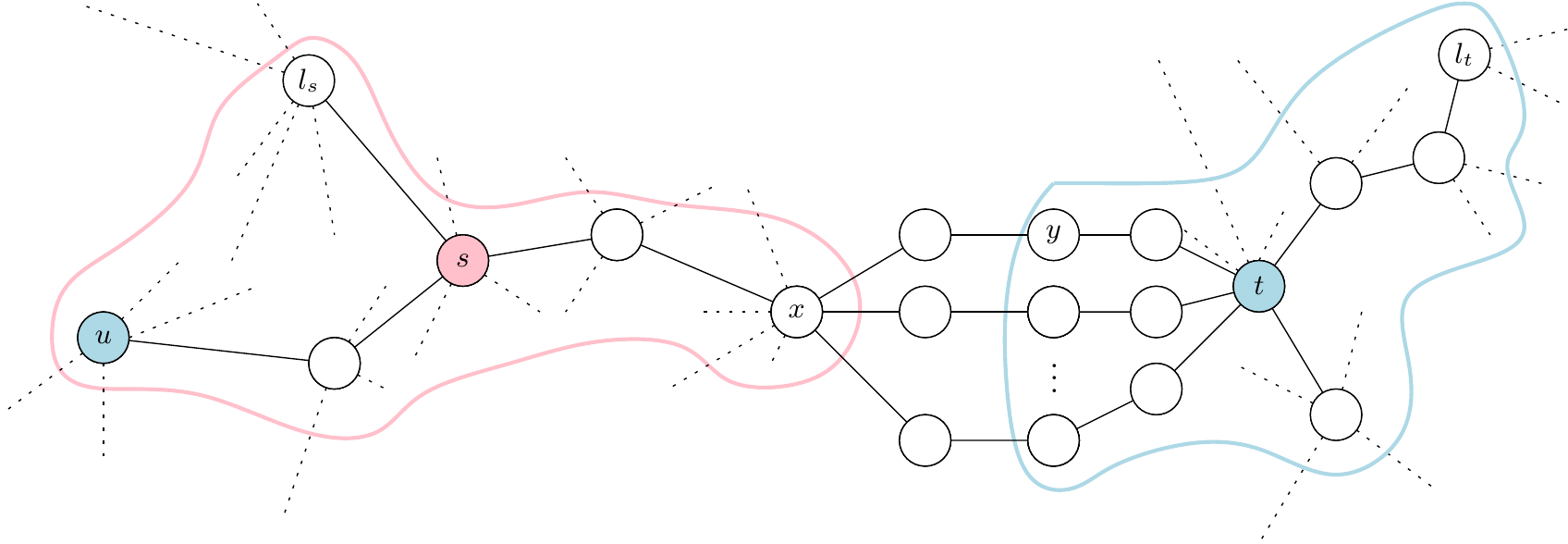}
	\centering
	\caption{There are nodes in a shortest path between $s$ and $t$ that do not belong to $\ball(s) \cup \ball(t)$.}
	\label{farBalls}
\end{figure}

We note that the routing scheme in~\cite{AGMNT04}, as sketched in the previous subsection, has some ``\emph{global}'' features that makes it plausibly difficult to locally verify, by adding certificates of size $\widetilde{O}(\sqrt{n})$ bits at each node. In this subsection, we mention one of these global features, illustrated in the example depicted on Figures~\ref{intersectBalls} and~\ref{farBalls}. 

In both Figures~\ref{intersectBalls} and~\ref{farBalls}, we are considering routing from a source~$s\notin L$ to a target~$t\notin L$, of different colors, with $s\notin\ball(t)$ and $t\notin \ball(s)$. Let us assume that node $s$ has color red, while node $t$ has color blue. According to the routing scheme in~\cite{AGMNT04}, node $s$ first routes the message towards some blue node $u\in\ball(s)$ to get information about the blue target~$t$. In the example of Fig.~\ref{intersectBalls}, node $u$ stores 
\[
\big(\rname_{T(u)}(s), x, port(x,y), \rname_{T(y)}(t)\big ),
\]
in order to guarantee a stretch~3. Instead, in the example of Fig.~\ref{farBalls}, node $u$ stores 
\[
\big ( \rname_{T(l_t)}(l_t), \rname_{T(l_t)}(t) \big ).
\]
to guarantee such a small stretch. Verifying locally whether there exists a good path between $s$ and $t$, which is the condition leading to distinguishing the case where the content of Eq.~\eqref{eq:caseA}  in Section~\ref{subsubsec:trt}  must be placed in the table, from the case where the content of Eq.~\eqref{eq:caseB2} must be placed in the table, appears to be a very difficult matter when restricted to certificates of size $\widetilde{O}(\sqrt{n})$.

\begin{figure}[htb]
	\includegraphics[width=0.5\textwidth]{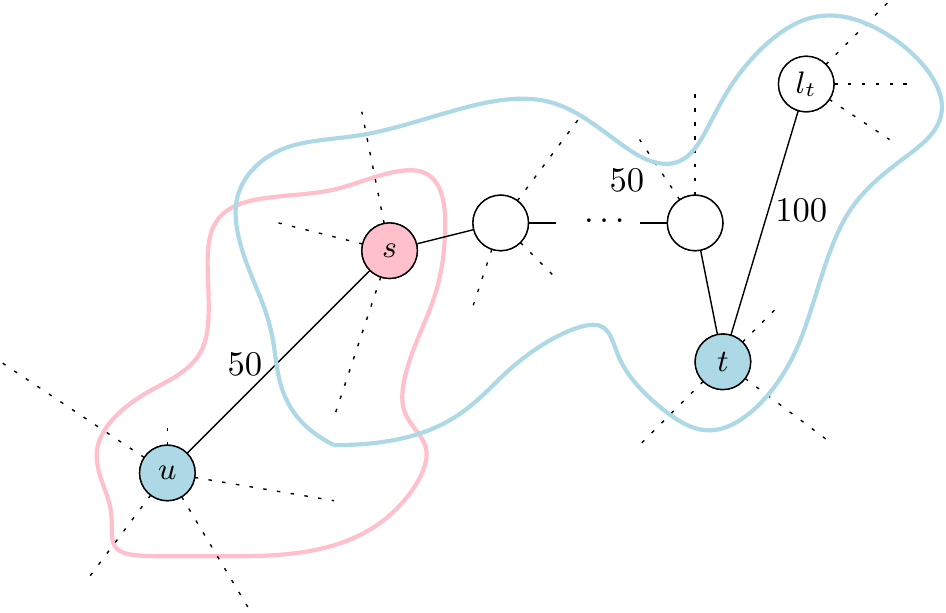}
	\centering
	\caption{A route of stretch~7.}
	\label{badExample}
\end{figure}

We point out that ignoring the two cases, and systematically storing the content of Eq.~\eqref{eq:caseA} would result in a routing scheme that may be locally verifiable. However, its stretch is at least $7$. To see why, let us consider the example displayed in Figure~\ref{badExample} where $s\in \ball(t)$ but $t \notin \ball(s)$. The radius of $\ball(s)$ is $50$. Let $u\in \ball(s)$ be one of the farthest nodes to $s$, i.e., $\delta(s,u)=50$. The radius of $\ball(t)$ is $100$ and $\delta(t,l_t)=100$. Although $\delta(u,t)=100$, we assume $u\notin \ball(t)$ due to  lexicographical order priorities. Finally, assume that $\delta(s,t)=50$. The worst case route from $s$ to $t$ would then be 
$
s \leadsto u \leadsto s \leadsto t \leadsto l_t \leadsto t.
$
This path is of length 

\begin{eqnarray*}
\ell &=& \delta(s,u) + \delta(u,s)  + \delta(s,t) + \delta(t,l_t) + \delta(l_t,t)\\ 
 &=& 2\times 50 + 50 + 2\times 100 = 350
\end{eqnarray*}

which is $7\delta(s,t)$. 

\break \section{A New Name-Independent Routing Scheme}
\label{sec:newnameind}

In this section, we describe and analyze a new name-independent routing scheme, denoted by $\cal R$.  The section is entirely dedicated to the proof of the following theorem. 

\begin{theorem}\label{theo:new}
	The name-independent routing scheme $\cal R$ uses routing tables of size $\widetilde{O}(\sqrt{n})$ bits at each node. It guarantees stretch~5, and, using handshaking, its stretch can be reduced to~3. In both cases, $\cal R$ can be locally verified using certificates of size $\widetilde{O}(\sqrt{n})$ bits, using a 1-sided error verification algorithm which guarantees that incorrect tables are detected with high probability\footnote{\small Given a family of events $({\cal E}_n)_{n\geq 1}$, event ${\cal E}_n$ holds with high probability if $\Pr[{\cal E}_n]=1-O(1/n^c)$ for $c\geq 1$.}.
\end{theorem}

Our verifier used to establish Theorem~\ref{theo:new} is actually deterministic. The certificates however store hash functions chosen at random. We assume that these hash functions are not corruptible, and that the adversary is not capable to create collisions (if not accidentally, by chance). Even though this assumption may seem strong, we point out that this is relevant to practical situations in which cryptographic hash functions designed to be collision resistant and hard to invert are used (for more details please refer to Chapter 5 of \cite{FSK10}).

\subsection{The new routing scheme $\cal{R}$}

Our new routing scheme $\cal{R}$ borrows ingredients from both \cite{AGMNT04} and \cite{TZ01}. In particular, the landmarks are chosen as for the (name-dependent) routing scheme in \cite{TZ01}, i.e., in such a way that every node $v$ has a cluster $\cluster(v)$ of size at most $4\sqrt{n}$, and the landmarks form a set $L$ of size at most $2\log n \sqrt{n}$. However, we reinforce the way the nearest landmark to $v$ is selected, by picking the nearest landmark $l_v$ such that 
\[
l_v=\argmin_{l \in L}\delta(v,l)
\]
where ties are broken by choosing the landmark with smallest identity. Also, we slightly reinforce the definition of $\rnext$: For every two nodes $v$ and $t$, we set 
\[
\rnext(v,t)
\]
as the \emph{smallest} port number at $v$ of an edge incident to $v$ on a shortest path between $v$ and $t$. These two reinforcements of the definitions of landmarks and $\rnext$ guarantee the following. 

\begin{lemma}\label{propLandmark}
	Let $v\in V$, and let $l_v$ be its  landmark. Let $u$ be a node on a shortest path between $v$ and $l_v$. Then $l_u=l_v$.
\end{lemma}

\begin{proof}
	By the choice of the landmarks, if $l_u\neq l_v$ then $\delta(u,l_u)<\delta(u,l_v)$. However, $\delta(v,l_u) \leq \delta(v,u)+ \delta(u,l_u)$. So, if $l_u\neq l_v$ then $\delta(v,l_u) < \delta(v,u)+ \delta(u,l_v) = \delta(v,l_v)$, contradicting the choice of $l_v$ as the  landmark closest to~$v$.  
\end{proof}

\begin{lemma}\label{propLandmark3}
	Let $v\in V$, let $l_v$ be its landmark, and let $w$ be the neighbor of $l_v$ such that $\rnext(l_v,v)=\rport_{l_v}(\{l_v,w)\}$. Let $u$ be a node on a shortest path between $v$ and $w$. We have $\rnext(l_v,u)=\rnext(l_v,v)$. 
\end{lemma}

\begin{proof}
	By Lemma~\ref{propLandmark}, we have $l_u=l_v$. If $\rnext(l_v,u)\neq\rnext(l_v,v)$ there  exists a shortest path $P$ from $l_v$ to $u$ that does not go to $w$ but through another neighbor $w'$ of $l_v$ with $\rnext(l_v,u)=\rport_{l_v}(\{l_v,w')\}$. 
	On the other hand, there is a shortest path between $l_v$ and $u$ going through $w$. So, $\rport_{l_v}(\{l_v,w'\})<\rport_{l_v}(\{l_v,w\})$. But the path consisting of $P$ and the shortest path between $u$ and $v$ is also a shortest path between $l_v$ and $v$. Therefore, by our reinforcement of the definition of $\rnext$, we get $\rport_{l_v}(\{l_v,w'\})>\rport_{l_v}(\{l_v,w\})$, contradiction. 
\end{proof}


As in \cite{AGMNT04}, each node that is not a landmark is given a color in $\{1,\dots,\sqrt{n}\}$ determined by a hash function, $\col$, that maps identities to colors, where at most $O(\log n \sqrt{n})$ nodes map to the same color. Also, each node $v$ has a vicinity ball, $\ball(v)$, that contains the  $O(\log(n) \sqrt{n})$ nodes closest to $v$ (breaking ties using identities). This guarantees that, with high probability, for every node $v$, there is at least one node of each color in $\ball(v)$. 

For each color $c$, where $1\le c\le \sqrt{n}$, we define a set  
\[
\Dir_c=\{ \big(v,l_v, \rnext(l_v,v) \big) : \col(v)=c \}
\]
which includes the direction to take at $l_v$ for reaching node~$v$ along a shortest path, for $v$ of color~$c$. 

\subsubsection{The routing tables}

Every node $v\in V$ then stores the following information in $\rtable(v)$: 

\begin{itemize}
	
	\item the hash function $\col$ that maps identities in colors;
	
	\item the identity of every landmark $l\in L$, and the corresponding port $\rnext(v,l)$;
	
	\item the identity of every node $u\in \cluster(v)$, and the corresponding port $\rnext(v,u)$;
	
	\item the identity of every node $u\in \ball(v)$, and the corresponding port $\rnext(v,u)$; 
	
	\item the set $\Dir_{\scol(v)}$. 
	
\end{itemize}

Storing the hash function $\col$ requires $O(\sqrt{n})$ bits, as we use the same function as in~\cite{AGMNT04}. Since $L$, $\cluster(v)$, $\ball(v)$, and $\Dir_{\scol(v)}$ are all of size $\widetilde{O}(\sqrt{n})$, we get that $|\rtable(v)|= \widetilde{O}(\sqrt{n})$ as desired. 

\subsubsection{Routing}

Let us consider routing towards a target node $t$, and let~$v$ be the current node. If $t\in\cluster(v)$, then routing to $t$ is achieved using $\rnext(v,t)$. Notice that, by Lemma~\ref{propCluster}, routing to $t$ will actually be achieved along a shortest path. Similarly, if $t\in\ball(v)$, then routing to $t$ is achieved using $\rnext(v,t)$ along a shortest path, and this also holds if $t$ is a landmark. In general, i.e., if $t$ is neither a landmark nor a node of $\cluster(v)\cup\ball(v)$, then node $v$ computes $\col(t)$ by hashing the identity of~$t$. 

If $\col(t)=\col(v)$, then $v$ forwards the message towards~$l_t$ using the information in $\Dir_{\scol(v)}$, and including $(l_t, \rnext(l_t,t))$ in the header of the message so that intermediate nodes carry on routing this message to $l_t$. At~$l_t$, the message will be routed to $t$ using the information $\rnext(l_t,t)$ available in the header, reaching a node $u_t$ such that $t\in\cluster(u_t)$. At this point, routing proceeds to $t$ along a shortest path. 

If $\col(t)\neq \col(v)$, then the message is forwarded to an arbitrary node $w \in \ball(v)$ having the same color as $t$ (we know that such a node exists), with $w$ in the header of the message. The message then reach $w$ along a shortest path. At $w$, we have $\col(t)=\col(w)$, and thus routing proceeds as in the previous case. 

\bigbreak

\noindent \textbf{Handshaking.} The routing with handshaking to node $t$ proceeds as follows. If $t\in L\cup \cluster(v)\cup\ball(v)$, or $\col(v)=\col(t)$, then routing proceeds as above. Otherwise, $v$ performs a handshake with a node $w \in \ball(v)$ with  $\col(w)=\col(t)$ in order to get the identity of $l_t$ as well as $\rnext(l_t,t)$. Then $v$ routes the message to $l_t$, where it is forwarded to a node $u_t$ such that $t\in\cluster(u_t)$. At this point, routing proceeds to $t$ along a shortest path.

\subsection{Stretch of the new routing scheme $\cal{R}$}

Let $s,t\in V$ be two arbitrary nodes of the graph. We show that the routing scheme $\cal R$ routes messages from $s$ to $t$ along a route of length at most $5\,\delta(s,t)$ in general, and along a route of length at most $3\,\delta(s,t)$ whenever using handshaking. 

As we already observed, if $t\in L\cup \cluster(v)\cup\ball(v)$, then the message is routed to $t$ along a shortest path, i.e., with stretch~1. Otherwise, we consider separately whether the color of $s$ is the same as the color of $t$, or not. 

Assume first that $\col(t)=\col(s)$. Then the message is routed towards $l_t$ along a shortest path, then from $l_t$ to $t$ along a shortest path. The length $\ell$ of this route satisfies
\[
\ell = \delta(s,l_t)+\delta(l_t,t).
\]
By the triangle inequality, we get that 
\[
\ell \leq \delta(s,t) + 2\delta(l_t,t).
\]
Since $t\notin \cluster(s)$, we get $\delta(s,t)\ge\delta(t,l_t)$. Therefore
\[
\ell \le 3\,\delta(s,t).
\]
We are left with the case where  $\col(s)\neq \col(t)$. Observe first that, with handshaking, the route from $s$ to $t$ will be exactly as the one  described in the case $\col(s)= \col(t)$, resulting in a stretch~3. This completes the proof that $\cal R$ achieves a stretch~3 with handshaking. Without handshaking,  the message is forwarded along a shortest path to an arbitrary node $w \in \ball(v)$ having the same color as $t$, then from $w$ to $l_t$ along a shortest path, and finally from $l_t$ to $t$ along a shortest path. The length $\ell$ of this route satisfies
\begin{eqnarray*}
	\ell & = &  \delta(s,w)+\delta(w,l_t)+\delta(l_t,t) \\
	& \leq & \delta(s,w)+\delta(w,s) +\delta(s,l_t)+\delta(l_t,t) \\
	& \leq & 2 \, \delta(s,w) + 3\,\delta(s,t) \\
	& \leq & 5\,\delta(s,t)
\end{eqnarray*}
as desired. 

\subsection{Local Verification of $\cal R$}

We show how to verify $\cal R$ with a verification algorithm $\verif$ using certificates on  $\widetilde{O}(\sqrt{n})$ bits. Let us define the certificates given to nodes when the routing tables are correctly set as specified by $\cal R$. 

For each color $c$, $1\le c\le \sqrt{n}$, let $B_c$ be the number of bits for encoding the set $\Dir_c$, and let $r=\max_{1\le c\le \sqrt{n}}B_c$. We have $r=\widetilde{O}(\sqrt{n})$. Let $f_1,\dots, f_k$ be $k = \Theta(\log n)$ hash functions, where each one is mapping sequences of at most $r$ bits onto a single bit. More specifically, each function $f_i$, $1\leq i \leq k$, is described as a sequence $f_{i,1},\dots,f_{i,r}$ of $r$  bits. Given a sequence $D=(d_1,\dots,d_\ell)$ of $\ell\leq r$ bits, we set 
\[
f_i(D) = \Big (\sum_{j=1}^{\ell} f_{i,j} \, d_j \Big ) \bmod 2. 
\]
Hence, if the $r$ bits describing $f_i$ are chosen independently uniformly at random, then, for every two $\ell$-bit sets $D$ and $D'$, we have~\cite{Yao79}: 
\[
D\neq D' \Rightarrow \Pr[f_i(D)=f_i(D')]=\frac 1 2
\]
Therefore, 
\[
D\neq D' \Rightarrow \Pr[(f_i(D)=f_i(D), \; i=1,\dots,k)]=1/2^k.
\]
That is, if $k=\beta\lceil \log_2n\rceil$ with $\beta>1$, applying the functions  $f_1,\dots, f_k$ to both sets $D$ and $D'$ enables to detect that they are distinct, with high probability $1-1/n^\beta$. 

Each node $v\in V$ stores the  certificate composed of the following fields: 

\begin{itemize} 
	
	\item the distances $\{\delta(v,l) : l\in L\}$;
	
	\item the distances $\{\delta(v,u) : u\in \ball(v)\}$;
	
	\item the set $\big \{\big(\delta(v,u),l_u,\delta(u,l_u)\big) : u\in \cluster(v)\big\}$;
	
	\item the set of $k$ hash functions $f_1,\dots, f_k$; 
	
	\item the set $\{f_i(\Dir_c) : 1\le i \le k, \; 1\le c \le \sqrt{n}\}$.
\end{itemize}

The first three entries are clearly on  $\widetilde{O}(\sqrt{n})$ bits, because of the sizes of $L$, $\cluster(v)$, and $\ball(v)$. Each function $f_i$ is described by a sequence of $\widetilde{O}(\sqrt{n})$ random bits.

\medskip

The verification algorithm $\verif$ then proceeds as follows. In a way similar to  the proof of Theorem~\ref{theo:classic}, we denote by $H(v)$, $L(v)$, $C(v)$, $B(v)$, $D(v)$, and $\{N(v,t) : t\in L(v) \cup C(v) \cup B(v)\}$ the content of the routing table of $v$. These entries are supposed to be the hash function $\col$, the set of landmarks, the cluster of $v$, the ball of $v$, the set $\Dir_{\scol(v)}$, and the set of $\rnext$-pointers given to~$v$, respectively. We also denote by $d$ the distance given in the certificates. That is, node $v$ is given a set $\{d(v,t) : t \in L(v)\cup B(v)\}$, and a set $\{(d(v,t),l_t,d(t,l_t)) : t\in C(v)\}$ where $l_t$ is supposed to be the node in $L(v)$ closest to $t\in C(v)$. We also denote by $F^v_1,\dots,F^v_k$ the hash functions given to $v$ in its certificate, and by
\[
F(v)=\{F^v_{i,c} : 1\le i \le  k, \; 1\le c \le \sqrt{n}\}
\]
the set of $O(\sqrt{n}\log n)$ hash values in the certificates. Of course, if a node does not have a table and a certificate of these forms, then it outputs \reject. So, we assume that all tables and certificates have the above formats. 

Clusters, balls and landmarks (including distances, ports, sizes, etc.) are checked exactly as in Section~\ref{sec:namedependent} for the routing scheme in~\cite{TZ01}. So,  in particular, ignoring the colors and ignoring the minimality of the landmarks' identities, we can assume that, for every node~$v$, 
\[
L(v) = L, \; C(v)=\cluster(v) \; \mbox{and} \; B(v) = \ball(v),
\]
and, for every node $u\in L \cup \cluster(v) \cup \ball(v)$, we have $d(v,u)=\delta(v,u)$, and $N(v,u)=\rnext(v,u)$ ignoring the minimality of that port number. To check the remaining entries in the routing tables (as well as the previously ignored colors and minimality criteria for the landmarks and the $\rnext$-pointers), $\verif$ performs the following sequence of steps. At every step, if the test is not passed at some node, then $\verif$ outputs \reject\/ at this node, and stops. Otherwise, it goes to the next step. If all tests are passed at a node, that node outputs \accept. Node $v$ checks that:

\begin{enumerate}
	\item \label{item:routingR1}  $l_v$ has the smallest identity among all landmarks closest to $v$; 
	
	\item \label{item:routingR2}  it has the same hash function $H(v)$ as its neighbors, that it has one node of each color in $B(v)$, that $v$ appears in $D(v)$, and that all nodes appearing in $D(v)$ have the same color as $v$; 
	
	\item \label{item:routingR3}  there exists at least one neighbor $u$ on a shortest path between $v$ and $l_v$ with  $N(l_v,v)=N(l_v,u)$;  for every neighbor $u$ on the shortest path between $v$ and $l_v$ (implying $l_u=l_v$) with $u\neq l_v$, $N(l_v,v)\leq N(l_u,u)$;  and if $l_v$ is a neighbor of $v$ on a shortest path between $l_v$ and $v$, $N(l_v,v)=\rport_{l_v}(\{l_v,v\})$; 
	
	\item \label{item:routingR4} for every $u\in L \cup B(v) \cup C(v)$, the port number $N(v,u)$ is the smallest among all the ports of edges incident to $v$ on a shortest path between $v$ and $u$; 
	
	\item \label{item:routingR5}  it has the same hash functions $F^v_1,\dots,F^v_k$, as its neighbors;
	
	\item \label{item:routingR6}  $F^v_i(D(v))=F^v_{i,H(v)}$ for every $i=1,\dots,k$;
	
	\item \label{item:routingR7}  the hash value $F^v_{i,c}$, $1\le i \le k$, $1\le c \le \sqrt{n}$, are identical to the ones of their neighbors.
\end{enumerate}

\noindent If $v$ passes all the above tests, then $v$ outputs \accept, else it outputs \reject. 

\medskip

We now establish the correctness of this local verification algorithm, that is, we show that it satisfies the specification stated in Section~\ref{subsec:certification}. First, by construction, if all tables are set according to the specification of $\cal R$, then every node running the verification algorithm with the appropriate certificate will face no inconsistencies with its neighbors, i.e., all the above tests are passed, leading every node to accept, as desired. So, it remains to show that if some tables are not what they should be according to~$\cal R$, then, no matter the certificates assigned to the nodes, at least one node will fail one of the tests with high probability. 

If all nodes pass the test of Step~\ref{item:routingR1}, then we are guaranteed that not only $L(v)=L$, but also that $l_v$ is indeed the appropriate landmark of~$v$. If all nodes pass the test of Step~\ref{item:routingR2}, then it must be the case that $H(v)=\col(v)$, that $B(v)=\ball(v)$ with the desired coloring property, and that 
\[
D(v)\subseteq \Dir_{\scol(v)}.
\] 
By Lemma~\ref{propLandmark3}, we get that if all nodes pass Steps~\ref{item:routingR3} and~\ref{item:routingR4}, then $N(v,u)=\rnext(v,u)$ where the minimality condition is satisfied. 

Let us assume that there exists a pair of nodes $(u,v)$ with same color such that 
\[
(u,l_u,\rnext(l_u,u)) \in \Dir_{\scol(v)}\setminus D(v). 
\]
By the previous steps, we know that $(u,l_u,\rnext(l_u,u))\in D(u)$.  Hence, $D(u)\neq D(v)$. On the other hand, if Step~\ref{item:routingR5} is passed by all nodes, then all nodes agree on a set of $k=\Theta(\log n)$ hash function $f_1,\dots,f_k$. Therefore,  assuming that these functions are set at random, we get that, with high probability, there exists at least one function $f_i$, $1\leq i \leq k$, such that $f_i(D(v))\neq f_i(D(u))$. 

If all nodes pass Step~\ref{item:routingR6}, then in particular $F^v_i(D(v))=F^v_{i,c}$, and $F^u_i(D(u))=F^u_{i,c}$ where $c=\col(v)=\col(u)$. We know that $F^v_{i,c}=f_i(D(v))$ and $F^u_{i,c}=f_i(D(u))$, which implies that $F^v_{i,c} \neq F^u_{i,c}$ with high probability, which will be detected at Step~\ref{item:routingR7} by two neighboring nodes in the network. This completes the proof of Theorem~\ref{theo:new}.

\break 
\section{Another name-independent Routing Scheme}
\label{sec:another}

We have seen that the stretch-3 name-independent routing scheme from~\cite{AGMNT04} does not seem to be locally checkable with certificates of $\widetilde{O}(\sqrt{n})$ bits. Our new name-independent routing scheme~$\cal R$ is locally checkable with certificates of $\widetilde{O}(\sqrt{n})$ bits, but it has stretch~5. In this section, we focus on the local checkability of the stretch-5 name-independent routing scheme from \cite{ACLRT03}. We show that, as for the scheme in ~\cite{AGMNT04}, the routing scheme in~\cite{ACLRT03} do not appear to be locally checkable with certificates of $\widetilde{O}(\sqrt{n})$ bits. Moreover, handshaking, which may allow that scheme to become locally verifiable with small certificates, does not reduce its stretch. 
\subsection{The Routing Scheme of~\cite{ACLRT03} }

The scheme in \cite{ACLRT03} requires $\widetilde{O}(\sqrt{n})$ bits of memory per node. It is based on the notion of landmarks, with as set $L$  of landmarks sith size $O(\log n \sqrt{n})$. To each node $v$ is also associated a vicinity ball, $\ball(v)$,  consisting of the $\sqrt{n}$ closest nodes to $v$, breaking ties using the lexicographic order of the identities. Each node has  at least one landmark in its vicinity ball. 

\subsubsection{Block partition}

Nodes are partitioned into $\sqrt{n}$ blocks, $S_1,\dots S_{\sqrt{n}}$, each containing $\sqrt{n}$ nodes. Assuming that the identities are from $0$ to $n-1$, the set $S_i$ consists of the nodes with identities in $[(i-1)\sqrt{n},\;i\sqrt{n})$. Set of blocks are assigned to the nodes. We denote by $A_v$ the set of blocks assigned to node $v$. We then denote by $R_v$  the set
\[
\{u\in V : \exists i \in \{1,\dots, \sqrt{n}\}, \; u\in S_i \; \mbox{and} \; S_i\in A_v\}.
\]
It is proved  in \cite{ACLRT03}  that there exists an assignment of sets of blocks to nodes such that 

\begin{itemize}
	\item for every $v \in V$, and for every block $S_i$, there exists $u\in \ball(v)$ such that $S_i\in A_u$;
	\item for every  $v\in V$, $|A_v|=O(\log n)$. 
\end{itemize}

\subsubsection{The routing tables}

This scheme  in \cite{ACLRT03} also uses routing schemes for trees. Similarly to the previously described  schemes, for a landmark $l\in L$, we denote by $T(l)$ a shortest-path tree routed at $l$ spanning all the nodes.  Also, $\rtable_{T}(v)$ and $\rname_T(v)$ denote  the routing table and name of node $v$, respectively, for the tree $T$. Each node $v\in V$ stores the following information in $\rtable(v)$: 

\begin{itemize}
	
	\item the identity of every node $u\in \ball(v)$, together with $\rnext(v,u)$; 
	
	\item the identity of every node $l\in L$,  together with $\rnext(v,l)$ and  $\rtable_{T(l)}(v)$;.
	
	\item for every node $u \in \ball(v)$, the set of indexes $i$ such that  $S_i\in A_u$; 
	
	\item the identity of every node $u\in R_v$ together with the identity of $l^*_{v,u}$, and $\rname_{T(l^*_{v,u})}(u)$ where 
	\[
	l^*_{v,u}=\argmin_{l\in L}\big(\delta(v,l) + \delta(l,u)\big).
	\] 
	
\end{itemize}

\subsubsection{Routing}

Routing from a node $s$ to a node $t$ is achieved in the following way. If $t\in \ball(s) \cup L \cup R_s$, then $s$ routes to $t$ using the information it has in its table. Otherwise, $s$ routes the message towards a node $w\in \ball(s)$ satisfying that $t\in R_w$. Then $w$ routes to $t$ using the information it has in its table, i.e.,  via the landmark $l^*_{w,t}$ that minimizes the sum $\delta(w,l) + \delta(l,t)$. It is proved in \cite{ACLRT03} that this routing scheme guarantees stretch~$5$.

\subsection{On the difficulty of locally verifying the scheme in \cite{ACLRT03}}

As for the routing scheme in~\cite{AGMNT04}, we note that the routing scheme in~\cite{ACLRT03} sketched in the previous subsection is plausibly difficult to locally verify by adding certificates of size $\widetilde{O}(\sqrt{n})$ bits at each node. Indeed, let us consider two nodes $u$ and $v$ with $u\in R_v$, one difficulty consists in verifying that a given landmark $l^*_{v,u}$ is effectively the one that, among all $l\in L$, minimizes the distance $\delta(v,l) + \delta(l,u)$. 

\begin{figure}[htb]
	\includegraphics[width=0.5\textwidth]{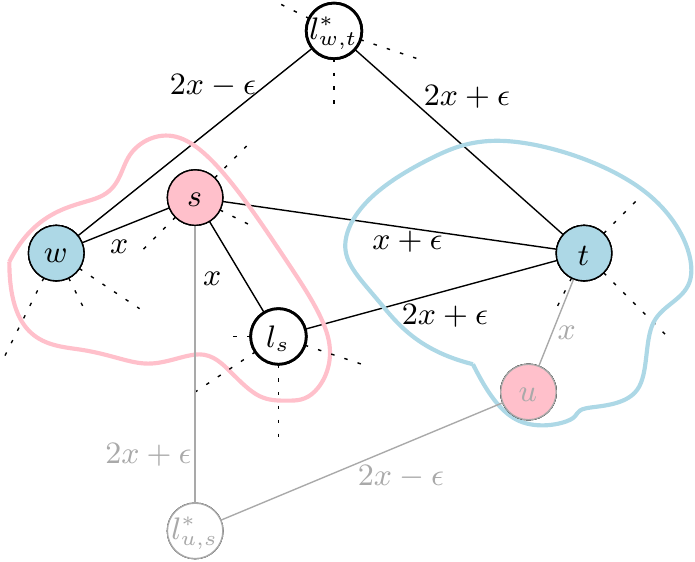}
	\centering
	\caption{Handshaking does not help.}
	\label{stretch5}
\end{figure}

Also, we show that, even using handshaking, the scheme in~\cite{ACLRT03} still achieves stretch at least~$5$. To see why, let us consider the example depicted in Fig.~\ref{stretch5}, where the node $s$ aims at routing towards the node~$t$. Node $w\in Ball(s)$ is the only node that has the routing information about node $t$, i.e., $t\in R_w$ and, for every $u\in \ball(s)\setminus \{w \}$, we have $t\notin R_u$.  Therefore, if the handshaking is performed to a node with information about~$t$ in $\ball(s)$, then this handshaking is to be performed between $s$ and $w$. Let us fix some positive values $x$ and $\epsilon$, with $\epsilon \ll x$. 

On Fig.~\ref{stretch5}, the following holds: 

\begin{itemize}
	
	\item $t\notin \ball(s)$, and $s\notin \ball(t)$;
	
	\item the radius of both $\ball(s)$ and $\ball(t)$ is $x$;
	
	\item $\delta(s,t)=x+\epsilon$;
	
	\item $\delta(s,w)=\delta(s,l_s)=x$ and $\delta(w,l_s)=2x$;
	
	\item $\delta(s,t)=x+\epsilon$;
	
	\item $ \delta(w,l^*_{w,t})=2x-\epsilon$ and $ \delta(l^*_{w,t},t)=2x+\epsilon$; 
	
	\item the shortest path between $s$ and $l^*$ goes through $w$, i.e., $\delta(s,l^*_{w,t}) = \delta(s,w) + \delta(w,l^*_{w,t})=3x-\epsilon$. 
\end{itemize}

Note that $\delta(l_s,t)=2x+\epsilon$, and thus 
\[
\delta(l_s,t)\le \delta(l_s,s) + \delta(s,t).
\]
Since $l^*_{w,t}=\argmin_{l\in L}\big(\delta(w,l) + \delta(l,t)\big)$,it follows that 
\[
\delta(w,l^*_{w,t}) + \delta(l^*_{w,t},t) \le \delta(w,l_s) + \delta(l_s,t).
\] 
In fact, in our example, we have  
\[
\delta(w,l^*_{w,t}) + \delta(l^*_{w,t},t) = 4x\le 4x + \epsilon =\delta(w,l_s) + \delta(l_s,t).
\]
Therefore, using handshaking, $s$ queries $w$ that replies with the name of $l^*_{w,t}$. Then $s$ sends the message to $t$ via $l^*_{w,t}$ and the resulting route is $s \leadsto w \leadsto l^*_{w,t} \leadsto t$, which has length:
\[
\begin{split}
\delta(s,w) + \delta(w,l^*_{w,t}) + \delta(l^*_{w,t},t) 
\\ = x + 2x - \epsilon + 2x + \epsilon = 5x 
\end{split}
\]
which is close to $5 \, \delta(s,t)$ for $\epsilon\ll x$. 

The case of handshaking with $t$ also leads to no improvements as far as stretch is concerned. See, e.g., Fig.~\ref{stretch5} where $t$ has a red node $u$ in its ball, at distance $x$ from $t$, and at distance $2x-\epsilon$ from $l^*_{u,s}$ with $l^*_{u,s}$ at distance $2x+\epsilon$ from $s$. This scenario leads to the very same situation as when the handshaking is performed between $s$ and $w$. 

\section{Generalization to larger stretches}
\label{app:generalization}

In this section, we show that Theorem~\ref{theo:classic} can be generalized to a whole family of routing schemes achieving a tradeoff between space complexity and stretch. More specifically, Thorup and Zwick describe in \cite{ThZw01} an algorithm which, given any integer $k\ge 1$, returns a collection $\cal T$ of trees such that every node is contained in $O(n^{1/k}\log^{1-1/k}n)$ trees in $\cal T$, and, for each pair of nodes, there exists a tree in $\cal T$ in which these two nodes are connected by a path of length at most $2k-1$ times the length of a shortest path between them. Thorup and Zwick use this result, combined with the techniques used in their stretch-$3$ routing scheme, to produce a family of routing schemes indexed by $k\geq 2$. For any $k\geq 2$ the routing scheme in this family uses a memory of $\widetilde{O}(n^{1/k})$ bits at each node, and has stretch $4k - 5$ (see \cite{TZ01}). The stretch can be reduced to $2k - 1$ using handshaking. 

\begin{theorem}\label{theo:general}
	For any $k\geq 2$, the stretch-$(4k-5)$ routing scheme by Thorup and Zwick in \cite{TZ01} can be locally verified using certificates of size $\widetilde{O}(n^{1/k})$ bits. This also holds for their handshaking-based stretch-$(2k-1)$ routing scheme. 
\end{theorem}

We provide below a sketch of the proof for Theorem~\ref{theo:general}. For this purpose, we recall the family of routing schemes in~\cite{TZ01}. 

\subsection{The Routing Scheme Family of \cite{TZ01}}

The design of this family combines techniques described in \cite{ThZw01} and \cite{TZ01}. 

\subsubsection{A hierarchy of landmarks}

The design is based on the construction of a hierarchy of landmarks 
\[
L_0\supseteq L_1 \supseteq \dots \supseteq L_{k-1} \supseteq L_k
\]
where $L_0=V$, $L_k=\emptyset$, and $|L_i|=\widetilde{O}(n^{1-i/k})$ for $i=0,\dots,k-1$. One way of constructing such set is to place each element of $L_{i-1}$ into $L_i$ independently with probability $n^{-1/k}$. This way, $L_{k-1}$ is of size $O(n^{1/k} \log n)$.  

For $0\le i\le k-1$, let $l_i(v)$ be the nearest landmark of node $v$ among all the landmarks in $L_i$. Notice that, for every node~$v$, we have  $l_0(v)=v$. Each node $v$ has a bunch associated to it, defined as follows.
\[
\bunch(v)=\bigcup_{0\le i < k}\{ u \in L_i\setminus L_{i+1} : \delta(v,u)< \delta (v,l_i(v)) \}.
\]
Note that, since $L_k=\emptyset$, we have $\delta(v,l_k(v))=+\infty$ for every $v\in V$, and thus, for every node $v$, 
\[
L_{k-1}\subseteq \bunch(v).
\] 
It can be shown~\cite{ThZw01} that, for any node $v$, 
\[
|\bunch(v)|=\widetilde{O}(n^{1/k}).
\]
The aforementioned collection $\cal T$ of trees is actually composed on $n$ trees, each rooted at a different node. We denote by $T(v)$ the tree in $\cal T$ that is rooted at~$v$. Moreover, $\cal T$ satisfies that, for every node $v$, and every $u\in \bunch(v)$, node $v$ belongs to $T(u)$. The same also hods for the landmarks, that is, for every node $v$, and every landmark $l_i(v)$, node $v$ belongs to $T(l_i(v))$. 

Each node has also a cluster associated to it, defined as follows.
\[
\cluster(v)=\bigcup_{0\le i < k}\{ u\in L_i\setminus L_{i+1} : \delta(v,u)< \delta (u,l_i(u)) \}.
\]
It can be shown~\cite{TZ01} that landmarks can be set in such a way that, for every $v$, 
\[
|\cluster(v)| = O(n^{1/k}).
\]

\subsubsection{Names and tables}

Each node $v$ stores in its table, $\rtable(v)$, the following information: 
\begin{itemize}
	
	\item the identities of each node $u\in \bunch(v)$;
	
	\item for every $u\in \bunch(v)$, the routing table $\rtable_{T(u)}(v)$ and the name $\rname_{T(u)}(v)$ of node $v$; 
	
	\item for every $u\in \cluster(v)$, the name $\rname_{T(v)}(u)$;
	
	\item the identity of the landmark $l_i(v)\in L_i$, $0\le i \le k-1$;
	
	\item the routing table $\rtable_{T(l_i(v))}(v)$.
\end{itemize}

All this information can be stored at $v$ using $\widetilde{O}(n^{1/k})$ bits. The $O(\log n)$-bit name of node $v$ is set as
\[
\rname(v)=\Big( \big (l_i(v), \rname_{T(l_i(v))}(v)\big), \; i=0,\dots,k-1 \Big). 
\]

\subsubsection{Routing}

The routing from a source $s$ to a target $t$ proceeds as follows. 

If $t\in \bunch(s)$, or $t=l_i(v)$ for some $i$, $0\le i \le k-1$, then $s$ routes to $t$ along a shortest path, using the information in its table, i.e., using $\rtable_{T(t)}(s)$. If $t\in \cluster(s)$, then $s$ routes  to $t$ along a shortest path using the information in its table, i.e.,using $\rname_{T(s)}(t)$ on the tree rooted at $s$. (Note that if $t\in \cluster(s)$, then $s\in \bunch(t)$, and, more generally,  $s\in \bunch(u)$ for any $u$ on a shortest path between $s$ and $t$, from which it follows that $u$ also knows how to route to $t$ using $\rtable_{T(s)}(u)$). 

Otherwise, node $s$ scans all identities $l_i(t)$ in $\rname(t)$, and, among all such landmarks in $\bunch(s)$, picks the one with the smallest index $i$, i.e., picks $l_i(t)$ where  
\[
i =\min\{j\in\{0,\dots,k-1\} : l_j(t)\in \bunch(s)\}.
\]
Then $s$  routes to $t$ using $\rname_{T(l_i(t))}(t)$ in the tree $T(l_i(t))$ --- this information can be extracted from the name of $t$. Note that there exists at least one landmark $l_i(t)\in \bunch(s)$ because $L_{k-1}\in \bunch(v)$ for every node $v$. 

It is proved in \cite{TZ01} that this routing scheme guarantees a stretch $4k-5$.

\medskip

\noindent \textbf{Handshaking.}  In order to reduce the stretch from $4k-5$ to $2k-1$, the source $s$ performs handshaking by querying node $t$, providing it with $\rname(s)$. The target  $t$ computes 
\[
j =\min\{\ell\in\{0,\dots,k-1\} : l_\ell(s)\in \bunch(t)\}.
\]
Then $t$ choses between $l_i(t)$ and $l_j(s)$ the one that has the minimum index among the two. That is, it picks 
\[
l^*=\left\{\begin{array}{ll}
l_i(t) & \mbox{if $i\leq j$} \\
l_j(s) & \mbox{otherwise}. 
\end{array} \right.
\]
and routing between $s$ and $t$ is performed along the tree $T(l^*)$, guaranteeing stretch $2k-1$.

\subsection{Sketch of the proof of Theorem~\ref{theo:general}}

It is not hard to check that one can locally verify the $O(\log^2n/\log\log n)$-bit routing tables in~\cite{FG01} for routing in trees with certificates of size $O(\log^2n/\log\log n)$ bits at each node. As for the basic case of stretch-3 routing with tables of size $\widetilde{O}(\sqrt{n})$ bits at each node, each node has to check the correctness of its bunch, its cluster,  and its set of landmarks. For this purpose, the certificate of every node~$v$ stores 

\begin{itemize}
	
	\item the distance $\delta(v,u)$ for every $u\in \{l_0(v),\dots,l_{k-1}(v)\}$, and every $u\in\cluster(v) \cup \bunch(v)$;
	
	\item the distance $\delta(u,l_u)$ for every node $u\in \cluster(v)$.
\end{itemize}

\noindent Each node $v$ then checks that:

\begin{enumerate} 
	
	\item its bunch, its cluster, and its set of the nearest landmarks are of the appropriate size;
	
	\item the set of landmarks $L_{k-1}$ is the same for every node;
	
	\item $l_i(v)$ is truly the nearest landmark of level $i$ of node $v$;
	
	\item all nodes in $\bunch(v)$ satisfy the definition of bunch, and that there are no missing nodes;
	
	\item the nodes in $\cluster(v)$ satisfy the definition of cluster, and that there are no missing nodes.
\end{enumerate}	

All these tests can be applied at each node by using the same techniques as to verify the stretch-$3$ routing scheme of \cite{TZ01}, and the correctness follows by the same arguments as those given in Section~\ref{sec:namedependent}.

\section{Conclusion and Further Work}
\label{sec:conclusion}

We have shown that it is possible to verify routing schemes based on tables of size $\widetilde{O}(\sqrt{n})$ bits using certificates with sizes of the same order of magnitude as the space consumed by the routing tables. The stretch factor is preserved, but to the cost of using handshaking mechanisms for name-independent routing. We do not know whether there exists a stretch-3 name-independent routing scheme, with tables of size $\widetilde{O}(\sqrt{n})$ bits, that can be verified using certificates on $\widetilde{O}(\sqrt{n})$ bits. Our new routing scheme, which is verifiable with certificates on $\widetilde{O}(\sqrt{n})$, has stretch~3 only if using handshaking (otherwise, it has stretch~5). Moreover, the certification of our routing scheme is probabilistic, and it would be of interest to figure out whether deterministic certification exists for some stretch-3 name-independent routing scheme, with tables and certificates on $\widetilde{O}(\sqrt{n})$ bits. It could also be of interest to figure out whether there exists a verification scheme using certificates of size $\widetilde{O}(n^c)$ bits, with $c<\frac12$. 

Interestingly, our result for stretch-3 name-dependent routing can be extended to larger stretches. Namely, by using the same techniques as for stretch~3, we can show that the family of routing schemes in~\cite{TZ01} using, for every $k\geq 1$, tables of size $\widetilde{O}(n^{1/k})$ with stretch at most $4k-5$ (or $2k-1$ using handshaking) are verifiable with certificates of size $\widetilde{O}(n^{1/k})$ (see \cite{BF17} for more details). However, we do not know whether such a tradeoff between table sizes and stretches can be established for verifiable \emph{name-independent} routing schemes. Specifically, are the existing families of name-independent routing schemes using, for every $k\geq 1$, tables of size $\widetilde{O}(n^{1/k})$ with stretch at most $O(k)$, verifiable with certificates of size $\widetilde{O}(n^{1/k})$? If not, is it possible to design a new family of verifiable name-independent routing schemes satisfying the same size-stretch tradeoff?

\bigskip

\noindent\textbf{Acknowledgements:} The authors are thankful to Cyril Gavoille and Laurent Viennot for discussions about the content of this paper. 



\end{document}